\documentstyle[prl,aps,multicol,epsf]{revtex}
\begin{document}
\title{Optimal orientation of striped states
in the quantum Hall system against external modulations}
\author{T. Aoyama, K. Ishikawa, and N. Maeda}
\address{
Department of Physics, Hokkaido University, 
Sapporo 060-0810, Japan}
\date{\today}
\maketitle
\begin{abstract}
We study striped states in the quantum Hall system around 
half-filled high Landau levels and
obtain the optimal orientation of the striped state in the presence 
of an external unidirectional periodic potential. 
It is shown that the optimal orientation is orthogonal to the 
external modulation in the Coulomb dominant regime (the orthogonal
 phase) and is parallel in the external modulation dominant regime (the
 parallel phase). 
The phase boundary of these two phases is determined numerically in the
 parameter space of the strength and wave number of the external
 modulation at the half-filled third Landau level. 
\end{abstract}
\draft

\begin{multicols}{2}
The modern semiconductor technology yields extremely pure
two-dimensional (2D) electron systems in heterostructures. 
In the presence of a strong perpendicular magnetic field, 
various kinds of new phenomena in addition to the ordinary quantum
Hall effect are found in the systems.
The effect of a crystal structure has been
believed to be ignored, because the magnetic length is much larger than
the lattice constant of the host crystal.
Hence, the system is supposed to have an orientational symmetry, 
that is, physics in the $x$ direction and $y$ direction is equivalent.  
In the end of the last century, however, highly anisotropic states,
which have an enormous anisotropy in magnetoresistances,
were observed around the half-filled
high Landau level (LL)\cite{Lilly,Du,Pan,Lilly2}.
This observation agrees with the striped state which was predicted 
in a mean field theory\cite{Shk,Moes}. 
The charge density of the striped state is uniform in the one direction and
periodic in the orthogonal direction.   
The Hartree-Fock (HF) theory and numerical calculations in a small system for 
striped states were studied recently at the 
filling factor where the anisotropy is observed 
and the results seem to support the striped
state\cite{Jung,Sta,Ma,Hal}. 
The experiments show that the stripe direction is parallel to the specific
crystallographic direction. 
The origin of the orientation of the striped state is still
puzzling\cite{Will,Coo}.

It is naturally considered that the origin of the orientation is related to 
yet unknown and very weak periodic structure in the sample. 
Therefore, 
we suppose that the origin to determine 
the orientation can be modeled
by an external modulation in the 2D electron system\cite{Ku}.
We compute the energy of striped states in the quantum Hall system
with an external modulation and find the optimal orientation.
It is naively expected that the external modulation breaks 
the orientational symmetry and makes the orientation of the striped state 
parallel to the modulation.
Unexpectedly, 
we find the counter-intuitive phase in which the optimal orientation of the 
striped state becomes orthogonal to the external modulation, and the other 
phase in which the optimal orientation is parallel to the external
modulation depending on the strength and wave number of the modulation. 
Our results are consistent with the recent experiments\cite{Will,Endo}.

Let us consider a 2D electron system in a perpendicular 
uniform magnetic field $B$ and a unidirectional periodic potential. 
The total Hamiltonian $H$ of the system is written as $H=H_0+H_1+H_2$, 
\begin{eqnarray}
H_0&=&\int \psi^\dagger({\bf r}){({\hat{\bf p}}+e{\bf A})^2\over2m}
\psi({\bf r})d^2r,\nonumber\\
H_1&=&{1\over2}\int\rho({\bf r})V({\bf r}-{\bf r}')\rho({\bf r}')d^2r
d^2r',\\
H_2&=&g\int\rho({\bf r})\cos ({\bf K}\cdot{\bf r})d^2r,\nonumber
\end{eqnarray}
where ${\hat p}_\alpha=-i\hbar\partial_\alpha$, 
$\partial_x A_y-\partial_y A_x=B$, 
$V=q^2/r$, $q^2=e^2/4\pi\epsilon$
($\epsilon$ is a dielectric constant).
$\psi({\bf r})$ is the electron field, 
and $\rho({\bf r})=\psi^\dagger({\bf r})\psi({\bf r})$. 
$H_0$ is the free Hamiltonian, which is quenched in the LL.
$H_1$ is the Coulomb interaction term and $H_2$ is the external modulation 
term. 
We ignore the spin degree of freedom.

The electron field is expanded by the momentum state $\vert f_l
\otimes\beta_{\bf p}\rangle$ in von Neumann lattice (vNL) 
formalism\cite{E} as 
\begin{equation}
\psi({\bf r})=\sum_{l=0}^{\infty}\int_{\rm BZ} {d^2 p\over(2\pi)^2}b_l
({\bf p})\langle{\bf r}\vert f_l\otimes\beta_{\bf p}\rangle,
\end{equation}
where $b_l({\bf p})$ is the anticommuting annihilation operator
with a LL index $l$ and a 2D lattice momentum
$\bf p$ defined in the Brillouin zone $\vert p_\alpha\vert\le\pi$. 
The annihilation operator $b_l({\bf p})$ obeys a twisted periodic boundary
condition $b_l({\bf p}-2 \pi {\bf N})
=e^{-i \pi (N_x+N_y)+i N_y p_x} b_l({\bf p})$, 
where $N_x$, $N_y$ are integers.
The momentum state is Fourier transform of the Wannier basis of 
vNL which localized at ${\bf r}=a (r_s m,n/r_s)$, 
where $n$, $m$ are integers. 
Here $a=\sqrt{2\pi\hbar/eB}$ and $r_s$ is an asymmetric parameter. 
We consider only the $l$ th LL state and ignore
the LL mixing. 
Hence, $H_0$ turns out to be constant and we omit the free Hamiltonian. 
Fourier transformed density operator ${\tilde\rho}$ in the $l$ th 
LL is written in vNL formalism as 
\begin{eqnarray}
{\tilde\rho}({\bf k})=\int_{\rm BZ}{d^2p\over (2\pi)^2}b_l^\dagger({\bf p})
b_l({\bf p}-a{\hat{\bf k}})f_l(k) \label{rho}\\
\times \exp[-i{a\over4\pi}{\hat k}_x(2p_y-a{\hat k}_y)],
\nonumber
\end{eqnarray}
where ${\hat{\bf k}}=(r_s k_x,k_y/r_s)$ and 
$f_l(k)=L_l({a^2k^2\over 4\pi})e^{-{a^2k^2\over8\pi}}$, 
here $L_l$ is the Laguerre polynomial. 
We substitute Eq.~(\ref{rho}) into $H_1$ and $H_2$ and find the ground 
state in two perturbative approaches. 
In the first approach, perturbative expansions with respect to $g$ 
in $H_2$, 
which describe the Coulomb dominant regime, are applied.
In the second approach, 
perturbative expansions with respect to $q^2/a$ in $H_1$, 
which describe the external modulation dominant regime, are applied. 
The filling factor is fixed at $l+1/2$ in the following calculation and
numerical calculations are performed at $l=2$. 

(I)
{\it $H_2$ as a perturbation} : We obtain the ground state of $H_1$ 
in the HF approximation first. 
Using the HF ground state, we treat $H_2$ as a perturbation.
This approximation is relevant in the Coulomb dominant regime, $g\ll q^2/a$.
We use the striped state  $\vert\Psi_1\rangle$ which is uniform 
in the $y$ direction as the unperturbed ground state of the $H_1$.
In the HF approximation, this is given as\cite{Imo,Ma}
\begin{equation}
\vert\Psi_1\rangle=N_1\prod_{\vert p_x\vert\le\pi,\vert p_y\vert\le\pi/2} 
b_l^\dagger({\bf p})\vert 0\rangle,
\label{eq:fermi}
\end{equation}
where $\vert0\rangle$ is the vacuum state for $b_l$ and $N_1$ is a 
normalization factor. 
The Fermi surface is parallel to the $p_x$ axis.
The density of this state $\langle\Psi_1\vert\rho({\bf
r})\vert\Psi_1\rangle$ is uniform in $y$ direction and periodic 
in $x$ direction with a period $ar_s$\cite{Ma,Imo}.
The orthogonality of the Fermi surface in the momentum space and the
density in the coordinate space plays important roles and is reminiscent
of the Hall effect. 
The one-particle spectrum is given by 
$\epsilon_{\rm HF}=\epsilon_{\rm H}+\epsilon_{\rm F}$,
\begin{eqnarray}
\epsilon_{\rm H}
&=&
\frac{2r_s q^2}{a\pi}\sum_{n=odd}
{f_l(2\pi n/ar_s)}^2
\frac{\cos(n p_y)}{n^2},
\\
\epsilon_{\rm F}
&=&
-\frac{r_s q^2}{a}\sum_{n=-\infty}^{\infty} \int_{-\frac{\pi}{2}+p_y+2\pi n}^{
\frac{\pi}{2}+p_y+2\pi n} dk_y
\int_{-\infty}^{\infty} \frac{dk_x}{2\pi}
h_l({\bf k})
,
\end{eqnarray}
where
$h_l({\bf k})={f_l(\sqrt{k_x^2+ (k_y r_s/a)^2 }) }^2/
\sqrt{k_x^2+(k_y  r_s/a)^2}$.
$\epsilon_{\rm HF}$ depends on only $p_y$, 
and the self-consistency condition for $\vert\Psi_1\rangle$ is satisfied.
The Fermi velocity is in the $y$ direction. 
The HF energy per particle is calculated as
$E_{\rm HF}(r_s)=\langle\Psi_1\vert H_1\vert \Psi_1\rangle/N$ where $N$ is
a number of electrons. 
$E_{\rm HF}$ is a function of $r_s$ and calculated as 
$E_{\rm HF}=\int_{-\pi/2}^{\pi/2} \frac{dp_y}{2\pi}
\epsilon_{\rm HF}(p_y)$ at the half-filled $l$ th LL. 
The optimal value $r_s=r_s^{\rm min}$ is determined so as to minimize 
$E_{\rm HF}(r_s)$\cite{Ma}.
The numerical value of $r_s^{\rm min}$ is $2.47$
at the half-filled third LL.

The perturbation energy per particle in the first order $\Delta E^{(1)}=
\langle\Psi_1\vert H_2\vert\Psi_1\rangle/N$ is written as 
\begin{equation}
\Delta E^{(1)}=\frac{g}{2N}
\langle\Psi_1\vert(\tilde{\rho}({\bf K})+\tilde{\rho}(-{\bf K}))
\vert\Psi_1\rangle. 
\end{equation}
The operator $\tilde{\rho}({\bf K})$ moves an electron in the Fermi sea 
by $a\hat{\bf K}$ in the momentum space. 
Therefore, except for the case that $a{\hat K}_y$ coincides with a 
multiple of $2\pi$, $\Delta E^{(1)}$ vanishes. 
We consider only the range $|a {\hat K}_y|< \pi$ which is sufficient to
compare our results with experiments. 

The perturbation energy per particle in the second order
$\Delta E^{(2)}$ is written as
\begin{equation}
\Delta E^{(2)} (g, \theta, K)=
\int_{\frac{\pi}{2}-a \hat{K}_y}^{\frac{\pi}{2}} \frac{dp_y}{2\pi}
\frac{-g^2 f_l(K)^2}{\epsilon_{\rm HF}(p_y+a \hat{K}_y)-\epsilon_{\rm HF}(p_y)},
\end{equation}
where ${\hat K}_y=K \sin{\theta}/r_s^{\rm min}$, $\theta$ is an angle 
between the stripe direction and external modulation.
We obtain the total energy per particle in the Coulomb dominant regime as 
\begin{equation}
E^{\rm Coul}(g, K, \theta)
=
E_{\rm HF}(r_s^{\rm min})+\Delta E^{(2)} (g, K, \theta).
\end{equation}
The $\theta$ dependence of $\Delta E^{(2)}$ is shown in Fig.~1 
at the half-filled third LL.
As seen in Fig. 1, the energy is always minimum at $\theta=\pi/2$, 
that is, the optimal orientation of the striped state is orthogonal to 
the external modulation. 
We call this phase the orthogonal phase.
Note that $\Delta E^{(2)}$ vanishes and $\theta$ dependence disappears
when $K$ equals the zeros of $f_l(K)$. 
In this case, 
the external modulation loses control of the stripe direction.

(II)
{\it $H_1$ as a perturbation} : We diagonalize $H_2$ 
first by choosing the $y$ axis of vNL to be parallel to the 
external modulation and $r_s=2\pi /aK$, that is, the period of the
striped states $ar_s$ equals the wave length $2\pi/K$ of
 the external modulation. 
Using this vNL basis, we treat $H_{\rm 1}$ as a perturbation. 
This approximation is relevant in the external modulation 
dominant regime, $g\gg q^2/a$.
Then the state $\vert \Psi_1\rangle$ is the ground state of $H_2$. 
The external modulation term reads 
\begin{equation}
H_2 = -\vert g  f_l( K) \vert \int_{\rm BZ} \frac{d^2 p}{(2\pi)^2}
b^\dagger_l({\bf p})b_l({\bf p}) \cos{p_y}.
\end{equation}
The ground state energy per particle of $H_2$ is obtained as 
$E_2 (g)= -\frac{2}{\pi} |g f_l(K)|$.
The perturbation energy per particle is calculated as
$E_{\rm HF}(\rm r_s)=\langle \Psi_1\vert H_1 \vert \Psi_1\rangle/N$ with 
$r_s=2\pi/aK$.
Hence,
the total energy per particle in the external modulation dominant regime
is given by  
\begin{equation}
E^{\rm ext}(g, K)=E_{\rm HF}(2\pi/aK)
-\frac{2}{\pi} \left \vert g f_l(K)
\right \vert.
\end{equation}
In this state, 
the stripe direction is parallel to the external modulation, 
which we call the parallel phase.

Finally we compare the total energies of striped states obtained in 
(I) and (II) at the half-filled third LL.  
As a typical example, 
$E^{\rm Coul}(g, K, \pi/2)$ and $E^{\rm ext}(g, K)$ for $aK=2$ 
is plotted in Fig. 2. 
The bold line represents the lower energy state.
As seen in Fig. 2, 
the orthogonal phase has lower energy and 
is realized at small $g$. 
The parallel phase has lower energy and is realized at large $g$. 
The phase boundary is calculated by solving the equation in
$g$, $E^{\rm ext}(g, K)=E^{\rm Coul}(g, K, \pi/2)$ for various value of
$aK$. 
The phase diagram in $g$-$K$ plane is shown in Fig. 3, 
where orthogonal and parallel phases are indicated by I and II respectively.
The dashed lines correspond to the zeros of $f_{2}(K)$, 
$aK=2.714$ and $6.550$, at which the stripe direction is undetermined. 
At $aK=2 \pi/r_{\rm min}=2.544$, where the period of the external
modulation coincides with the optimal period of the stripe, 
the phase boundary touches the $K$ axis.

The direct verification of our results is made by observing a
transition between the two phases by tuning the extenal modulation.
The necessary wave length of the modulation for the verification
is on the order of $2a$, which is about 100nm at $B=2 {\rm T}$. 
Recently, the unidirectional lateral superlattice with a period 92nm
is achieved on top of the 2D electron system\cite{Endo}.
The experiment shows that the magnetoresistance orthogonal to the external
modulation has a shallow and broad dent between 
two peaks around $\nu=9/2$.
The magnetoresistance parallel to the external modulation does not have
the same structure around $\nu=9/2$.
Anisotropy observed in this experiment is small due to the low
mobility compared with experiments of striped states
\cite{Lilly,Du,Pan,Lilly2}. 
The strength of the modulation is estimated as $g=0.015{\rm meV}$. 
The parameters $(g, K)=(0.006 q^2/a, 3.097 /a)$ correspond to this
experimental setting, and are shown as $X$ in Fig.~4. 
$X$ belongs to the orthogonal phase.
In this phase, the one-particle dispersion has no energy gap in the orthogonal
 direction 
and has an energy gap in the parallel direction to the external modulation.
Hence, the magnetoresistance orthogonal to the external modulation is
strongly modified by the injected electric current compared with 
the parallel magnetoresistance. 
This is consistent with the experiment. 
With a slightly larger period 115nm, 
orthogonal magnetoresistance around $\nu=9/2$ is structureless. 
In this case, 
the corresponding parameters $(g, K)=(0.014 q^2/a, 2.478 /a)$ are
 shown as $Y$ in Fig.~4 \cite{private}. 
$Y$ belongs to the parallel phase. 
Hence, there is an energy gap in orthogonal direction to the external
modulation and the magnetoresistance in this direction is not 
modified strongly by the injected electric current.
This is also consistent with the experiment.
We hope that a similar experiment with a higher mobility sample will give
more clear evidence for our results.

In the half-filled lowest LL, 
the anisotropic effect is observed under the external modulation
\cite{Willet2}.
The semiclassical composite fermion theory is proposed for this 
experiment\cite{Oppen}. 
In this theory, 
it is assumed that anisotropic transport is caused by the density modulation. 
On the other hand, we study the striped 
state under the external modulation in the half-filled higher LL.
Note that the origin of the anisotropy in the present case is
 spontaneous stripe formation rather than the external modulation. 

It seems difficult to understand how the orthogonal phase is realized
contrary to the naive expectation that two striped structures tend to be 
parallel.
To understand the reason, it is convenient to consider the striped state in the
momentum space. 
Since the Fermi surface of the striped state is flat as seen in
Eq.~(\ref{eq:fermi}), a perturbation with a wave number vector 
perpendicular to the Fermi surface affects the total energy most strongly. 
Therefore the orthogonal phase could be realized in a small external
modulation. 
The point of our theory is that the mean field theory has the flat Fermi 
surface. 
The fluctuation around the mean field has been studied but 
discussions seem unsettled yet\cite{Fra,Mac,Cote}. 
Comparisons between experiments with an in-plane magnetic
field\cite{Pan,Lilly2,Pan2} and
HF calculations indicate that the mean field energy is good
approximation for the striped state\cite{Jung,Sta}. 
Higher order corrections are expected to be small 
because the Fermi velocity diverges due to the Coulomb interaction\cite{Imo}. 
We estimate $\Delta E^{(2)}$ in the RPA approximation to the density 
correlation function as $\pi^{\rm{RPA}}_{00}(k)=\pi_{00}(k)/(1-\tilde{V}(k) \pi_{00}(k))$.
The results are shown in Fig. 1 by dashed lines. As seen in this figure
 the correction is small actually.

In summary it is shown that a weak external modulation determines the
orientation of the striped state and 
there are two phases in the 2D 
parameter space of the strength and wave number of the external modulation, 
that is, the orthogonal phase and parallel phase.
In the former phase, the optimal orientation of the striped state is 
orthogonal to the external modulation. 
In the latter phase, the optimal orientation is 
parallel to the external modulation. 
The phase diagram is obtained numerically at the half-filled third 
LL. 
We believe that our findings shed a new light on the origin of an orientation
of striped states in quantum Hall systems.

We thank A. Endo and Y. Iye for useful discussions. 
This work was partially supported by the special Grant-in-Aid for
 Promotion of Education and Science in Hokkaido University provided by
 the Ministry of Education,
Science, Sports, and Culture, and by the Grant-in-Aid for Scientific
Research on Priority area (Physics of CP violation) (Grant No. 12014201).

\begin{figure}
\centerline{
\epsfxsize=3in \epsffile{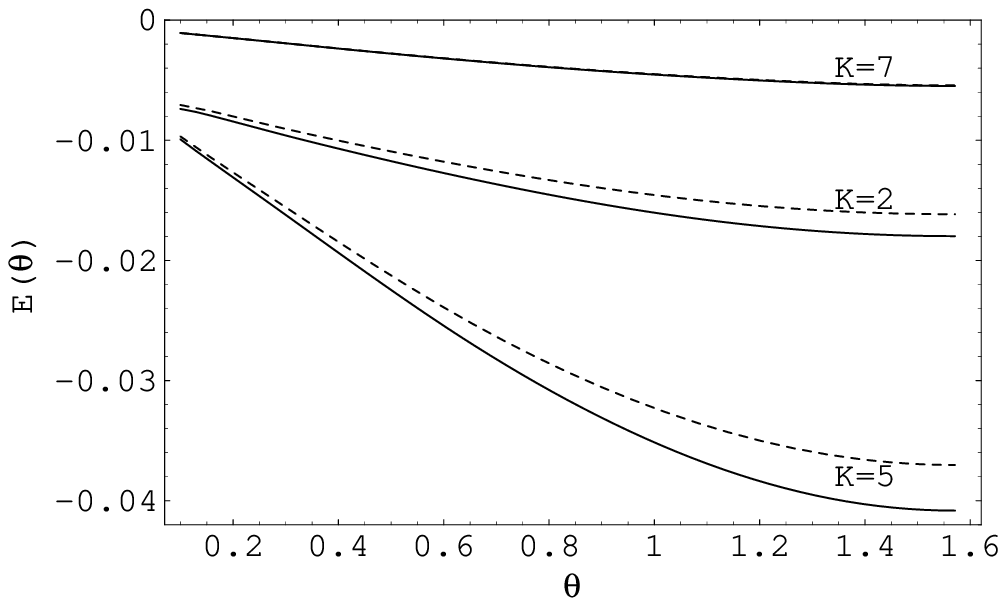}}
Fig. 1. The $\theta$-dependence of $\Delta E^{(2)}(g, K, \theta)$ 
for $aK=2, 5, 7$. The solid lines and dashed lines stand for the HF 
calculation and RPA approximation, respectively.
The unit of energy is $g^2 a/q^2$
\label{fig:phase}
\end{figure}
\begin{figure}
\centerline{
\epsfxsize=2.7in \epsffile{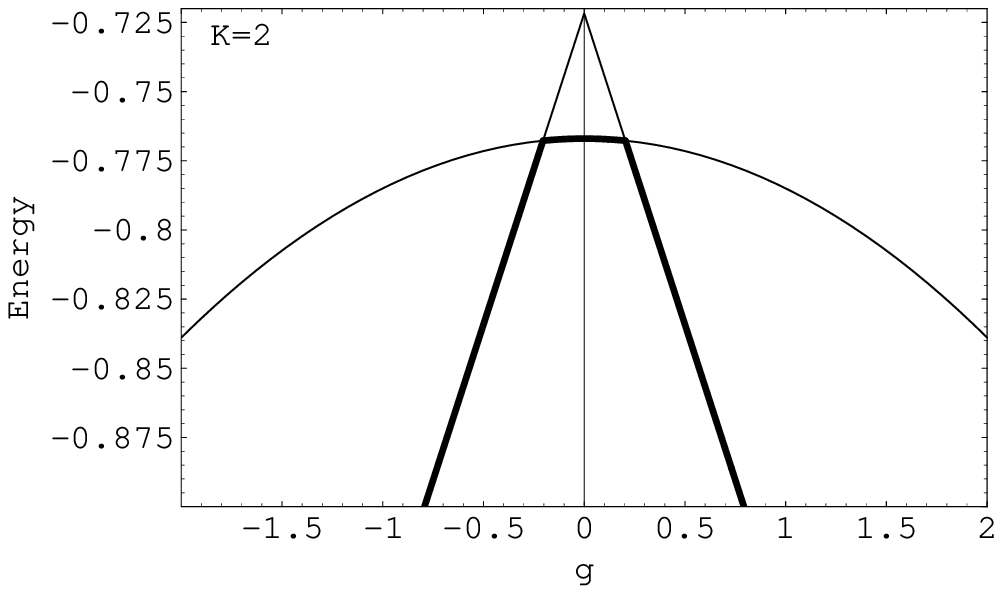}}
Fig. 2. The total energy of the striped state obtained in (I) and (II)
 for $aK=2$. 
The unit of energy and $g$ is $q^2/a$. 
The straight line stands for $E^{\rm ext}(g, 2/a)$ and the parabolic line
 stands for $E^{\rm Coul}(g, 2/a, \pi/2)$.
The lower energy state is represented by the bold line.
\label{fig:phase1}
\end{figure}
\begin{figure}
\centerline{
\epsfxsize=3in \epsffile{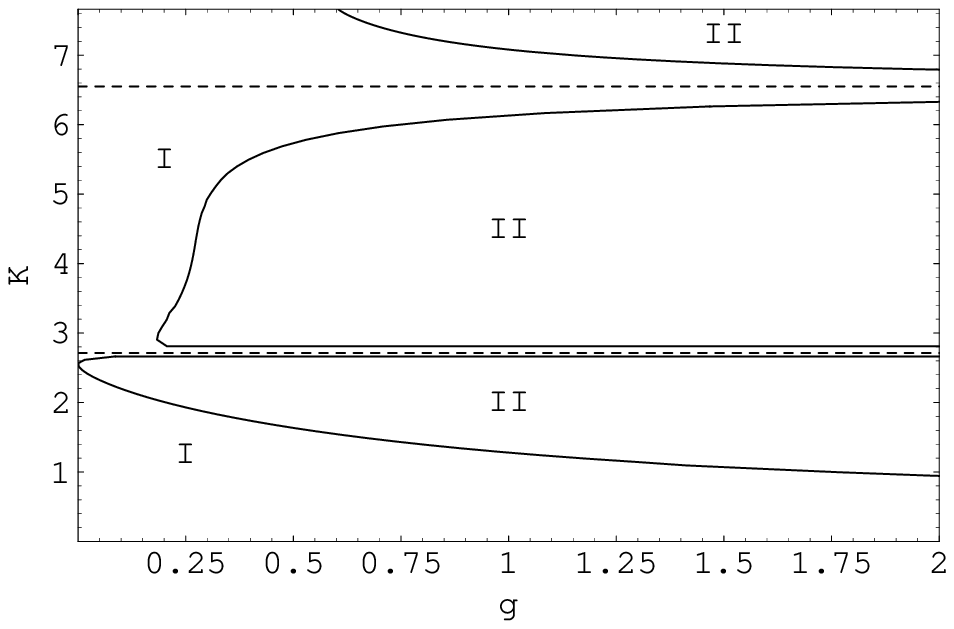}}
Fig. 3. The phase diagram of the striped state at the half-filled third 
 LL.
The unit of $g$ is $q^2/a$ and the unit of $K$ is $1/a$. 
The regions denoted by I and II correspond to the orthogonal phase and
parallel phase, respectively. 
The dashed lines represent the zeros of $f_{2}(K)$, 
at which the stripe direction is undetermined.
\label{fig:phase2}
\end{figure}
\begin{figure}
\centerline{\epsfxsize=2.7in \epsffile{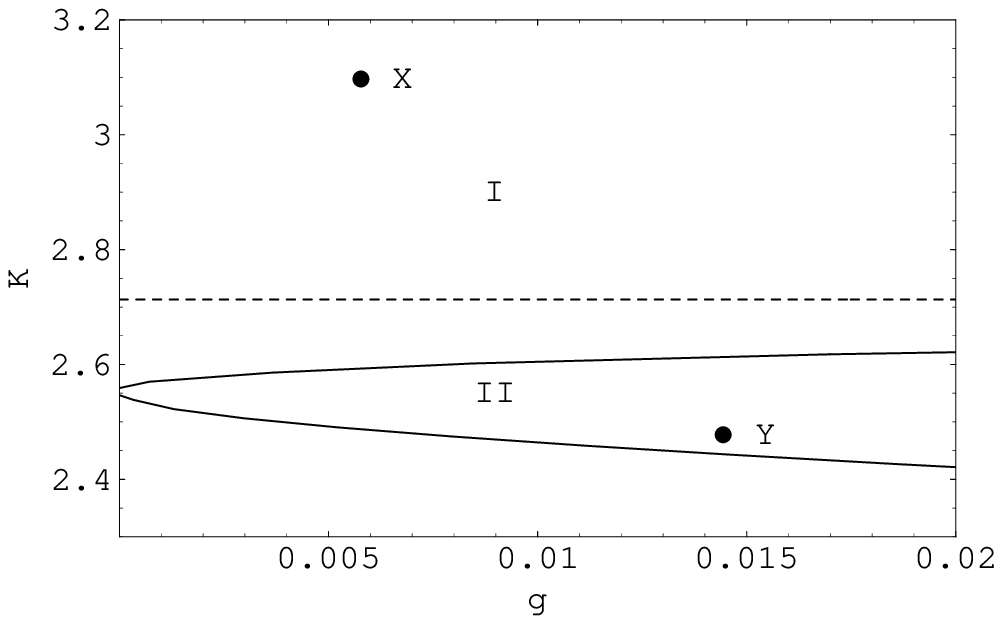}}
Fig. 4. The two points $X$ and $Y$ in the phase diagram
stand for the experimantal data\cite{Endo}.
\label{fig:phase2}
\end{figure}

\end{multicols}
\end{document}